\documentclass[amssymb,amsmath,aps,showpacs,floatfix,nofootinbib,showpacs,12pt]{revtex4}
\usepackage{graphicx,epsfig}
\def\beq{\begin{equation}}
\def\eeq{\end{equation}}
\def\be{\begin{equation}}
\def\ee{\end{equation}}
\def\bea{\begin{eqnarray}}
\def\eea{\end{eqnarray}}
\def\nnb{\nonumber}

\newcommand{\gsim}{\lower.7ex\hbox{$\;\stackrel{\textstyle>}{\sim}\;$}}
\newcommand{\lsim}{\lower.7ex\hbox{$\;\stackrel{\textstyle<}{\sim}\;$}}

\begin{document}

%\newcommand{\Romannum4}{\uppercase\expandafter{\romannumeral4}} 
%\newcommand{\rmnum1}{\romannumeral1}
%\newcommand{\rmnum4}{\romannumeral4}

%\begin{center}
 %\vspace{0.2cm}
 \title{ Charged lepton flavor violation on target at GeV scale}
 \author{ Wei Liao and Xiao-Hong Wu}
 \affiliation{
  Institute of Modern Physics, School of Sciences \\
 East China University of Science and Technology, \\
 130 Meilong Road, Shanghai 200237, P.R. China %
}

%\end{center}

\begin{abstract}
We study the lepton flavor violating process, $e+ T \to \tau +T'$,  at a few GeV. 
This process can be studied by experiments directing GeV scale electron or positron 
beams on internal or fixed targets. We study the effects of some low energy lepton 
flavor violating interactions on this process. We study the sensitivities of this process 
on these low energy lepton flavor violating interactions and compare them to the sensitivities
of lepton flavor violating $\tau$ decay processes. Comparing with $\tau$ decay processes,
this process provides another way to study the lepton flavor violating effects with $e-\tau$
conversion and it can be searched for in facilities with GeV scale electron or positron beams 
which are available in a number of laboratories in the world.

\end{abstract}
 \maketitle

 {\bf Introduction}

As very important possible evidence towards the physics beyond the Standard Model(SM),  
processes of charged lepton flavor violation(cLFV) have attracted a lot of attention.  
This is partly because flavor violation of neutral leptons has been observed in oscillation of neutrinos.
It raises the hope that some flavor violating effects of charged leptons, although small but not undetectable,  might also appear,
if the dynamics underlying the neutrino flavor does not treat the flavor of charged leptons badly unfair.
Flavor violating decay processes of  muon, such as $\mu \to e \gamma$, $\mu \to 3e $ , have been intensively studied by researchers
and have been probed to high precision,  thanks to the intense muon source available in experiment.
They are expected to be further probed to a  very high precision by future experiment  with muon source
improved by orders of magnitude.

Flavor violation related to $\tau$ lepton has not been investigated with fever as for muon. 
Current upper bound to processes $\tau \to e \gamma$, $\tau \to  3e$, $\tau \to \mu \gamma$, $\tau \to \mu e^+ e^-$ etc. ,
are many orders of magnitude weaker than those similar processes of  $\mu$ decay.
The major difference is that $\tau$ source is much harder to prepare than the $\mu$ source,
because of its larger mass and much smaller lifetime.  
If high luminosity $\tau$-charm factory or Z factory could be built, these $\tau$ decay processes can be investigated to much higher precision.
Before these possible facilities, we are also allowed to probe lepton flavor violating effect associated with $\tau$ lepton
by considering inverse processes with electron or positron beam on target.
The experiment can be done with electron or positron beam directed on a fixed target or internal target.
The typical processes to observe are $e^\pm +T \to \tau^\pm + T'$ where $T$ is the target nucleon or nucleus and $T'$ represent all
final particles in the nuclear part of the process.  This kind experiment can be done for electron/positron beam with
several GeV. Such kind of GeV scale electron/positron beams  are available at facilities such as BEPCII,  SLAC,  JLAB etc. 

In this article we are going to investigate this kind of cLFV processes with electron or positron
on target and study the sensitivity of these kind of processes on the lepton flavor violating interaction 
associated with $\tau$ lepton. In the next section we first discuss the general kinematics of the process under discussion.
Then we concentrate on several processes with typical final states $T'$.  
One process to discuss is the elastic scattering with $T'=T$.  
The other one is the quasi-elastic scattering in which a nucleon is kicked out of nucleus by elastic scattering with lepton.
We discuss the sensitivity of these two typical processes on two types of lepton flavor violating
interactions, namely the $e \tau \gamma$ vertex and $e \tau Z$ vertex.
Throughout this article we concentrate on $e-\tau$ conversion using electron or positron beam.
One can simply apply the discussion in this article to $\mu-\tau$ conversion for appropriate
muon beam if it's available.
 
 \vskip 0.5cm
 {\bf  Kinematics in scattering of $e-\tau$ conversion}
 
 We consider production of $\tau$ lepton in scattering of electron or positron with target $T$
 \bea
 e^\mp(k) +T(P) \to \tau^\mp(k') +T' (P'), \label{scattering0}
 \eea
where $k$,$k'$,$P$ are four-momenta of corresponding particles and  $P'$ is the total momentum of the final
product, $T'$, which represents all final particles in the target part of the process. 
The initial target $T$ can be nucleon or some heavy nuclei.
In general, $T'$ can be a complicated product.  We are not going to discuss very complicated cases in later sections.
Instead, we concentrate on some simple processes with $e-\tau$ conversion: 1) elastic scattering(ES)  with $T'=T$;  
2) quasi-elastic(QE) scattering in which a nucleon is kicked out of nucleus by elastic scattering with $e^\mp$,
so that T breaks into several pieces  but without a change in the number of nucleon. 
Two other important processes for scattering at a few GeV scale are 1) in-elastic scattering with $T'$ being the excited state of $T$;
2) process with pion production.  
These two processes are more or less related to the topics in this article but will not be discussed in detail.

Introducing  $q=P'-P=k-k'$ and $Q^2=-q^2$, we can find that
\bea
Q^2=-(k-k')^2=2k\cdot k'-m_\tau^2, \label{kinematics01}
\eea
where electron mass has been neglected  in comparison with the appearance of $m_\tau$,  the mass of $\tau$ lepton.
Throughout this article, we always neglect the electron mass in our result.
For target $T$ at rest, we have $P\cdot q=m_T(E-E')=m_T(E_{T'}-E_T)$.  $m_T$ is the
mass of the target T.  $E$, $E'$,
$E_T$ and $E_{T'}$ are the energies of $e$, $\tau$, T and T$^\prime$  separately.

From $P'^2=(q+P)^2$ we can get
\bea
Q^2 &&=-q^2=2P\cdot q-(m^2_{T^\prime}-m^2_T)  \nnb \\
&&=2 m_T(E-E')-(m^2_{T^\prime}-m^2_T).
% =2m_T(E_{T'}-E_T)-(m^2_{T^\prime}-m^2_T). 
\label{kinematics02}
\eea
where $m_{T'}$ is the invariant mass of $T'$: $m^2_{T'}=P'^2$.

\begin{table}
\begin{tabular}{|c|c|c|c|c|c|}
 \hline
 Target  & proton & deteuron & Helium-3 & Helium-4 & Lithium-7\cr
 \hline
 Threshold E(GeV) & 3.44 & 2.61 & 2.33 & 2.19 & 2.01 \cr
 \hline
 \end{tabular}
 \caption{Threshold energy of $e +T \to \tau +T'$ process for a few target nuclei.} 
 \label{Threshold}
 \end{table}

Using the scattering angle $\theta_{k'}$, the angle between the direction of ${\vec k}'$  and the direction of 
the incoming electron or positron beam,  $E'$ can be solved as illustrated in Appendix A. 
From (\ref{kinem10}) one can see that for $e +T \to \tau +T'$  process to happen the following condition should be satisfied
\bea 
(m^2_\tau-m^2_{T'}+m^2_T+2 m_T E)^2-  4 m_\tau^2  ( 2 E m_T+m^2_T) >0.\label{kinematics03}
\eea
One can find from (\ref{kinematics03}) that the initial energy should satisfy
\bea
E > m_\tau \frac{m_{T'}}{m_T} +\frac{1}{2 m_T} (m^2_\tau +m^2_{T'}-m^2_T). \label{kinematics04}
\eea

For elastic scattering,  we have $T'=T$ and $m_{T'}=m_T$.  We can find
\bea
Q^2  =2 m_T(E-E'),
% =2m_T(E_{T'}-E_T)-(m^2_{T^\prime}-m^2_T). 
\label{kinematics02a}
\eea
and the threshold condition becomes
\bea
E > m_\tau  +\frac{m^2_\tau}{2 m_T} . \label{kinematics05}
\eea
Apparently condition (\ref{kinematics05}) is weaker than (\ref{kinematics04}), and is the threshold energy
condition for this lepton flavor violating process. 
In Table \ref{Threshold} we can see numerically, with a few possible target, 
this requirement on the beam energy  for  experiments with this lepton flavor violating process.

 \begin{figure}[tb]
\begin{center}
\begin{tabular}{cc}
\includegraphics[scale=1,width=8cm]{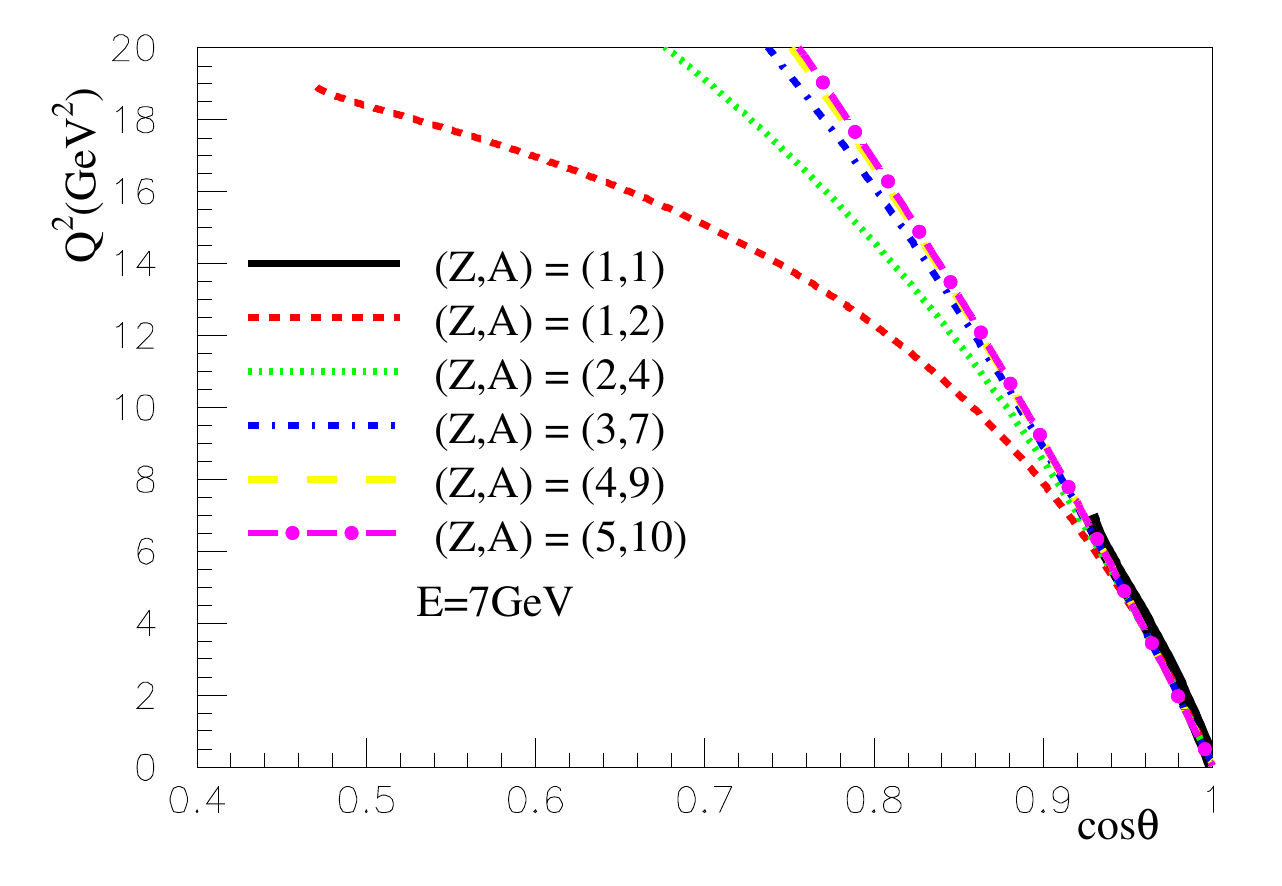}
\includegraphics[scale=1,width=8cm]{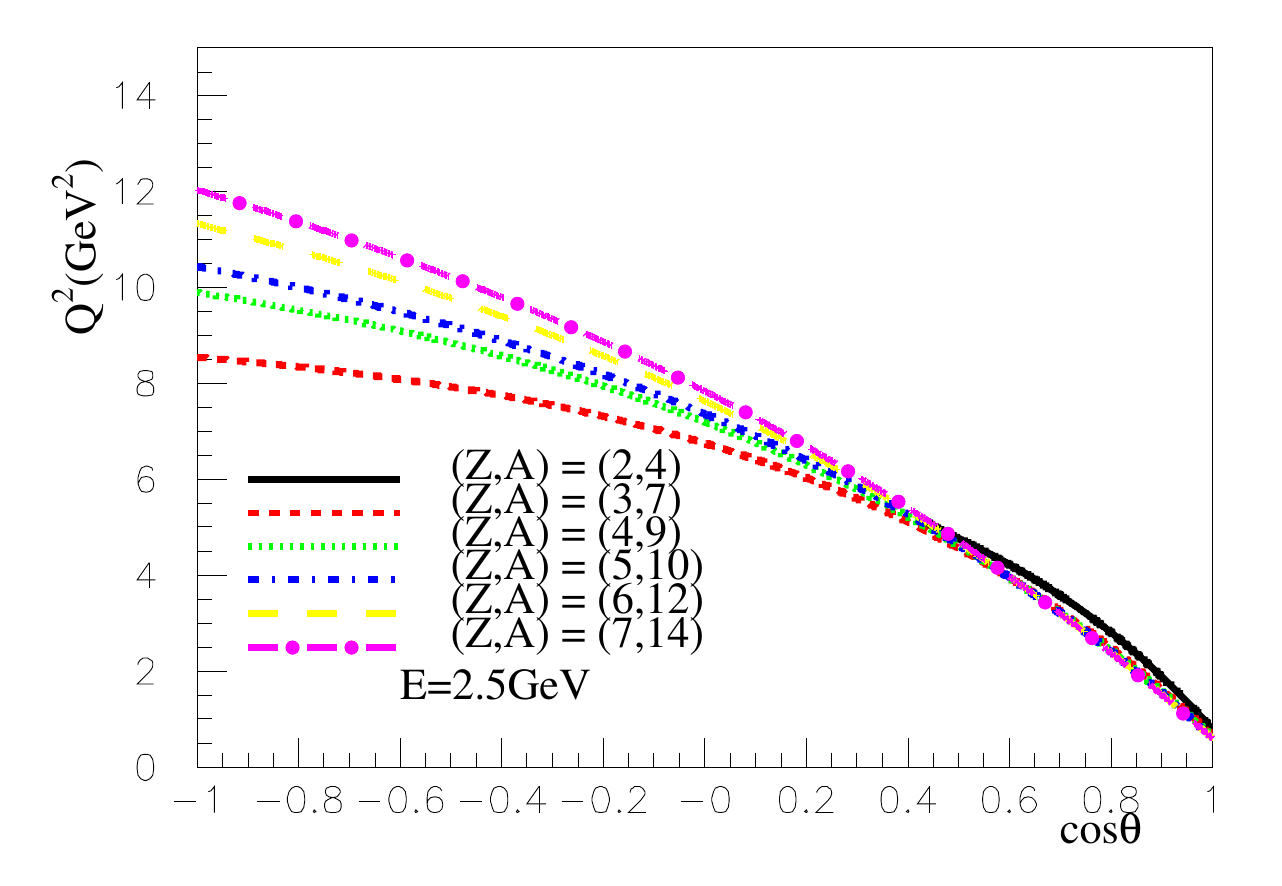}
%\\
%\includegraphics[scale=1,width=7.5cm]{figure1c}
%\includegraphics[scale=1,width=7.5cm]{figure1d}
%\\
%\includegraphics[scale=1,width=7.5cm]{figure1e}
%\includegraphics[scale=1,width=7.5cm]{figure1f}
\end{tabular}
\end{center}
\caption{$Q^2$ versus $\cos\theta_{k'}$ for  elastic scattering $e +T \to \tau +T$.}
\label{Q2vstheta}
\end{figure}

 \begin{figure}[tb]
\begin{center}
\begin{tabular}{cc}
\includegraphics[scale=1,width=8cm]{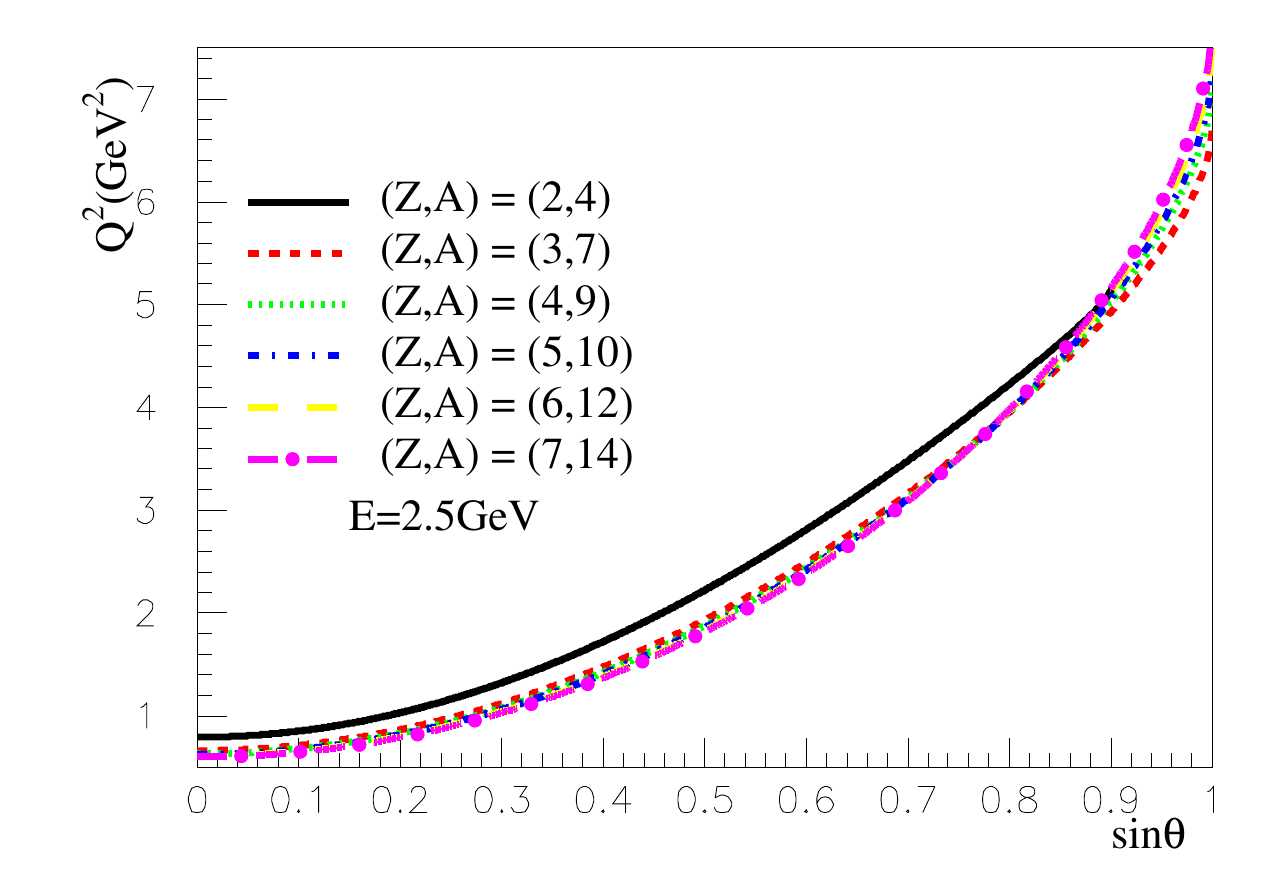}
\includegraphics[scale=1,width=8cm]{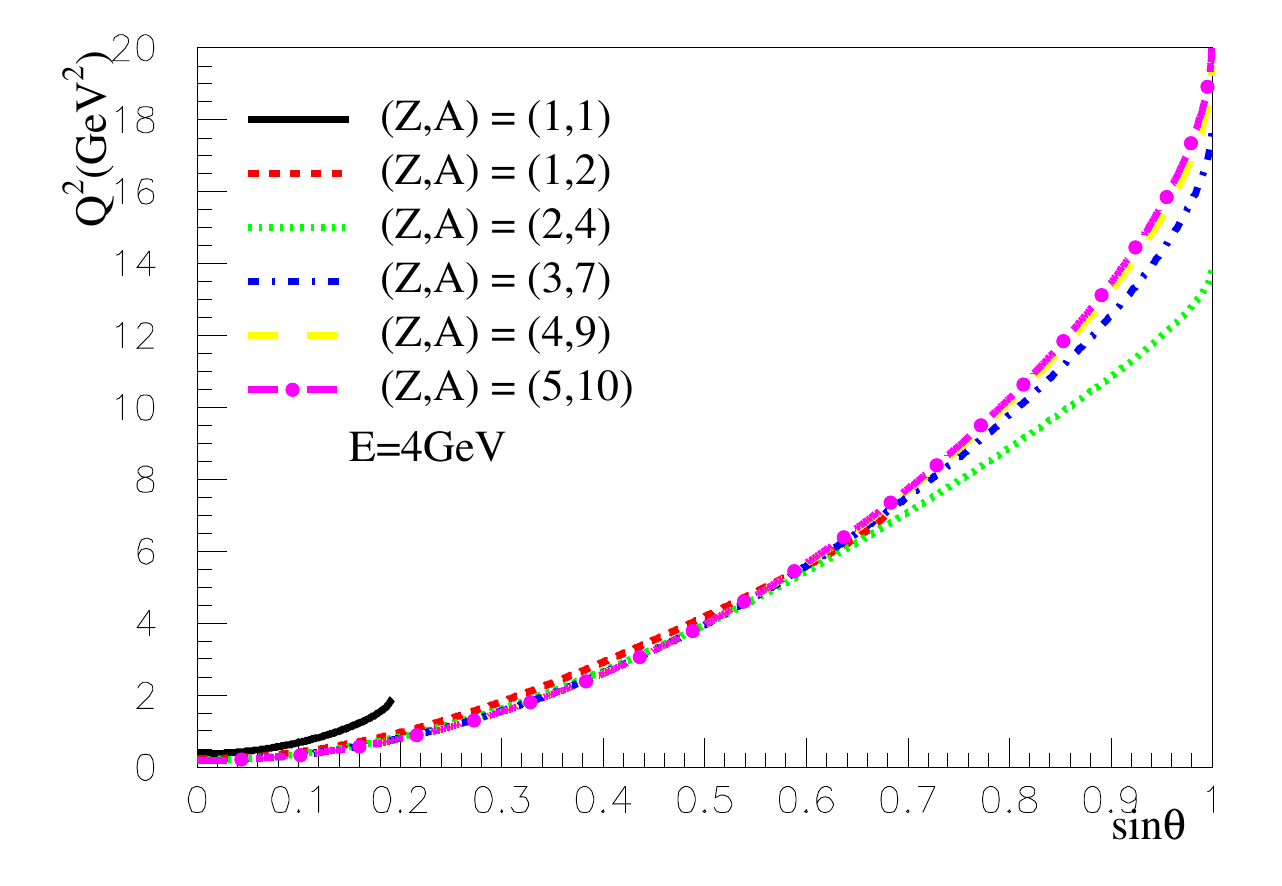}
\\
\includegraphics[scale=1,width=8cm]{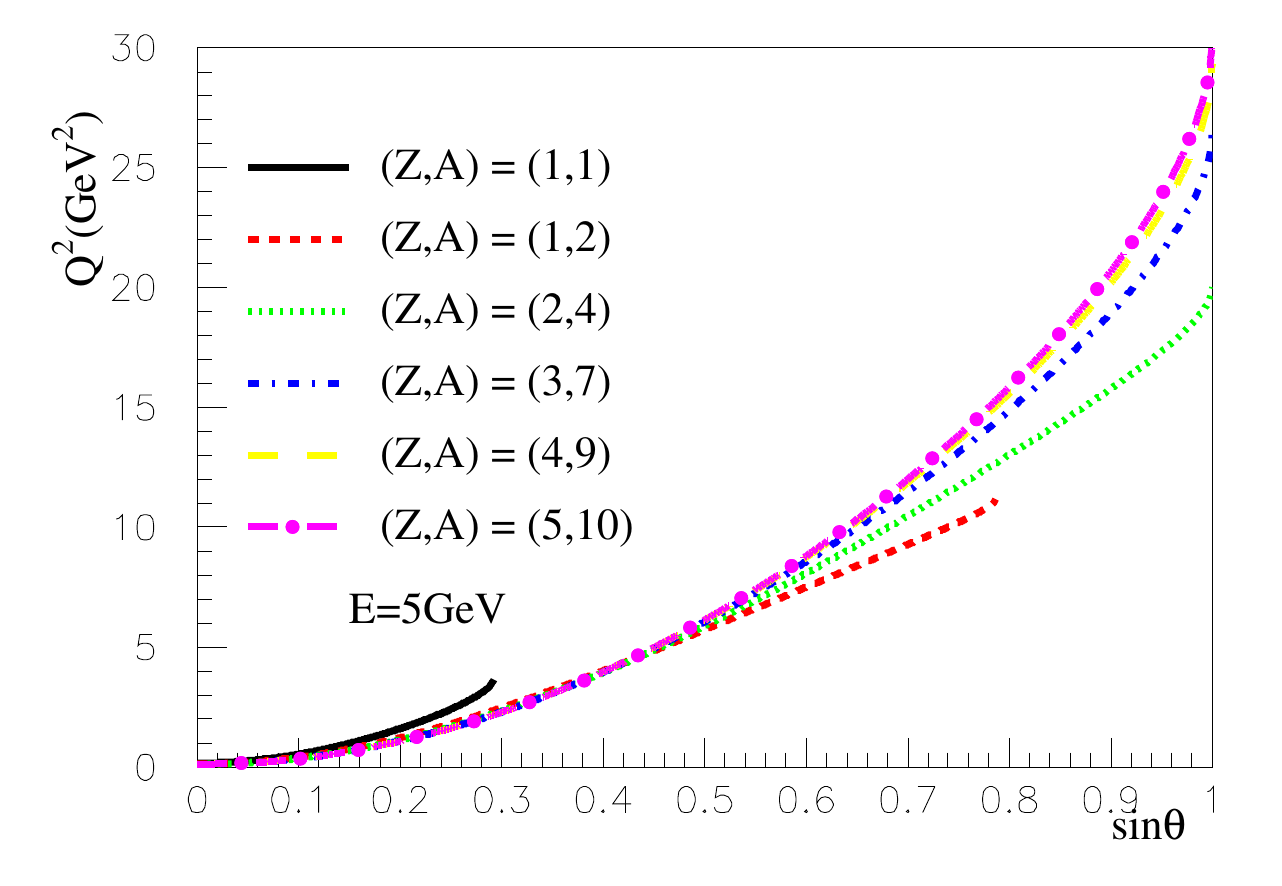}
\includegraphics[scale=1,width=8cm]{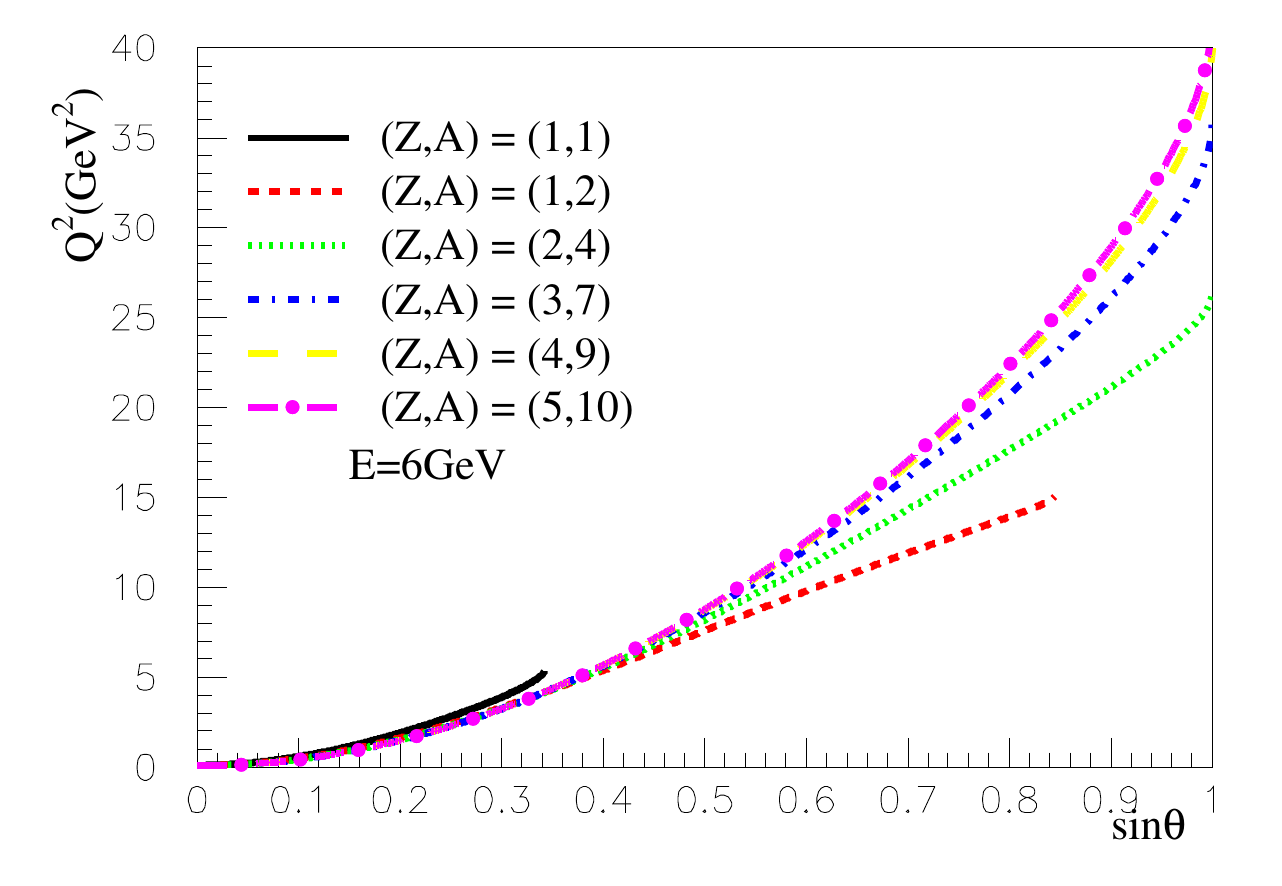}
%\\
%\includegraphics[scale=1,width=7.5cm]{figure1e}
%\includegraphics[scale=1,width=7.5cm]{figure1f}
\end{tabular}
\end{center}
\caption{$Q^2$ versus $\sin\theta_{k'}$ for  elastic scattering $e +T \to \tau +T$.}
\label{Q2vstheta2}
\end{figure}

$Q^2$ versus the scattering angle $\theta_{k'}$ in the elastic case  can be obtained  using (\ref{kinematics02a}) and
the solution of $E'$ in (\ref{kinem03}). 
In Fig.  \ref{Q2vstheta}  we show this dependence numerically for $E=2.5$ GeV and $E=7$ GeV
respectively.  One can see that the scattered direction of $\tau$ lepton is limited to a range 
close to the forward direction due to a kinematic constraint, as discussed in detail in Appendix A,  namely  (\ref{kinem05}). 
One can see in Fig. \ref{Q2vstheta2} that this range of $\theta_{k'}$, although not very large, can still reach
 $\sin\theta_{k'}\approx 0.2-0.3$ for $E=4-6$ GeV which is fairly large. So there are rooms detecting
the final $\tau$ lepton even for a detector having problems to detect final particles moving 
close to the forward direction.

One thing to notice in this process is that due to the unavoidable momentum loss to target in $e \to \tau$ conversion, a minimal amount
of energy has to be transferred to the nuclear sector and $Q^2$ has nonzero minimum in the scattering.
In Table \ref{Q2min} one can see $Q^2_{min}$, the minimum of $Q^2$,  for various nuclei and
beam energies.   The larger the energy or heavier the target nucleus,  the smaller the $Q^2_{min}$ can reach. 
For a fixed E,  $Q^2_{min}$ approaches a constant as the target mass increases, but does not approaches to zero.
For larger energy, the situation would approach to the case in which massless limit applies. That is why $Q^2_{min}$
turns to be closer to zero for larger E.

\begin{table}
\begin{tabular}{|c|c|c|c|c|c|c|}
 \hline
 Target(Z,A)  & (1,1)  & (1,2) & (2,4) & (3,7) & (4,9) & (5,10) \cr
 \hline
 Q$^2_{min}$(GeV$^2$) (E=$2.5$ GeV) &  $\times$ &   $\times$ &  0.79 & 0.66  & 0.63 & 0.62 \cr
  \hline
 Q$^2_{min}$(GeV$^2$) (E=$4$ GeV) &  0.37 &  0.23 &  0.20 & 0.19  & 0.18 & 0.18 \cr
 \hline
 Q$^2_{min}$(GeV$^2$) (E=$5$ GeV) & 0.17 &  0.13  &  0.12  &  0.11  &  0.11 & 0.11 \cr
 \hline
 Q$^2_{min}$(GeV$^2$) (E=$6$ GeV) &  0.10 &  0.085 &  0.078 & 0.076  & 0.075 & 0.075 \cr
 \hline
Q$^2_{min}$(GeV$^2$) (E=$7$ GeV) &  0.071   & 0.060  & 0.056  & 0.055 &  0.054 &  0.054 \cr
  \hline
 \end{tabular}
 \caption{Minimal $Q^2$ for elastic scattering $e +T \to \tau +T$. } 
 \label{Q2min}
 \end{table}

In quasi-elastic scattering, the electron/positron is considered to scatter elastically with a nucleon in a nucleus
and kicks this nucleon out of the target nucleus. The target nucleus can break into several pieces due to this scattering,
but there is no extra nucleon or other hadron produced in this process.  If the energy and energy transfer are large enough,  
the scattering cross section of this QE process can be computed using the impulse approximation in which 
the initial nucleon is considered at rest inside the nucleus, and outgoing nucleon and the incoming and outgoing leptons are 
all approximated as plane waves.  This approximation greatly simplifies the calculation of this process.
In this article we adopt this approximation for QE scattering.

For in-elastic scattering, i.e. when $T'$ is an excited state of nucleus, the situation is a bit complicated than
ES or QE processes. The discussion would depend on the details of the excited states of various nuclei
and results associated with them  have to be discussed specifically. Moreover, it depends on the detail of
the detector in experiment, i.e.  whether a specific excited state of a nucleus could be detected.
More importantly,  we expect that using in-elastic scattering would not improve very much the sensitivity to  
$e-\tau$ lepton flavor violating interactions.
So we are not going to elaborate them in the present article.

Production of pion or other heavier particles  in the neutral current processes with energy of a few GeV and  $Q^2$ of a few GeV$^2$
are usually not dominant,
comparing with the neutral current elastic scattering with nucleon $e +N \to e+ N$.
So,  considering the processes with production of pion or other hadron with $e-\tau$ lepton flavor conversion is not going
to improve a lot the sensitivity on  $e-\tau$ lepton flavor violating interactions 
discussed in this article .
So we are not going to elaborate on these kind processes and leave more detailed researches on them to
future works.

\vskip 0.5cm
 {\bf Probing $e \tau \gamma$  interaction}
 
 As an important example of lepton flavor violating interaction,  $e-\tau$ conversion induced by the interaction
 with a photon should be studied.
 In this section we study the  sensitivity of lepton flavor violating ES and QE scattering processes
 on the $e \tau \gamma$ interaction vertex.
  
 The effective Lagrangian governing the $e\tau \gamma$  interaction vertex can be written as
\bea
\Delta L=e  c_L {\bar \tau}_R i\sigma^{\mu \nu} e_L F_{\mu \nu}+
e c_R {\bar \tau}_L i\sigma^{\mu \nu} e_R F_{\mu \nu} +h.c. \label{ETauGammaV}
\eea
where $c_{L,R}$ are coupling strengths with dimension -1 for left-handed and right-handed electron field operators respectively.
$F_{\mu \nu}$ is the electromagnetic field strength tensor.
From (\ref{ETauGammaV}) we can get the branching ratio of the $\tau \to e \gamma$ decay
\bea
Br(\tau \to e \gamma)=\tau_\tau \alpha (|c_L|^2+|c_R|^2) m^3_\tau , \label{Decayrate2}
\eea
where $\tau_\tau$ is the lifetime of $\tau$ and $\alpha$  the fine structure constant. 
For the present upper bound Br$(\tau \to e \gamma) < 3.3\times 10^{-8}$ ~\cite{pdg} we have 
\bea
|c_L|^2+|c_R|^2 < 1.8\times 10^{-18} ~\textrm{GeV}^{-2} \label{constraint2}
\eea

For one photon exchange processes, one can write the cross section of $e + T \to \tau +T'$ scattering as follows
\bea
\frac{d \sigma}{d Q^2}=\frac{\pi Z^2 \alpha^2}{ Q^4 E^2} W_{\mu \nu} L^{\mu \nu} \label{Xsec2}
\eea
where $W_{\mu \nu}$ is the nuclear(hadronic) tensor and  $L^{\mu \nu}$ the leptonic tensor.
The nuclear tensor can be parametrized in general as \cite{structure, drell, eTscattering}
\bea
&& W_{\mu \nu} =-(\eta_{\mu \nu} - q_\mu q_\nu/q^2)  W_1 \nnb \\
&&~~~+\frac{1}{m_T^2} (P_\mu -q_\mu P\cdot q/q^2)
(P_\nu -q_\nu P\cdot q/q^2) W_2.\label{Ntensor2}
\eea
The nuclear tensor obeys $W_{\mu \nu} q^\nu =q^\mu W_{\mu \nu}=0$ which results from
the current conservation. 
The nuclear tensor is common to one photon exchange processes in which final states in leptonic part can be different 
but final $T'$ is the same. So $W_{\mu \nu}$  can be measured in electron scattering processes $e +T \to e+ T'$. 

For $e+T\to e+T'$ scattering, the leptonic tensor is the familiar form
\bea
L^{\mu \nu}= 2(k^\mu k'^\nu+k^\nu k'^\mu -k\cdot k' \eta^{\mu \nu}) \label{Ltensor2a}
\eea
For $e+T \to \tau+ T'$  scattering, we can get from (\ref{ETauGammaV}) the leptonic tensor as
\bea
L^{\mu \nu}&&=-2(|c_L|^2+|c_R|^2)[m_\tau^2(m_\tau^2-q^2)(\eta^{\mu \nu}-q^\mu q^\nu/q^2) \nnb \\
&&  +4 q^2(k^\mu -q^\mu k\cdot q/q^2)(k^\nu-q^\nu k\cdot q/q^2) ], \label{Ltensor2b}
\eea
where $q=k-k'$ and $q^2=-Q^2$ as given in the last section. The anti-symmetric part of the leptonic tensor
does not contribute to the final result and has been neglected in (\ref{Ltensor2b}).

For elastic scattering in which $T'=T$, the nuclear tensor can be written as
\bea
&& \textrm{spin 0} :  ~W_1=0,~W_2=|F(Q^2)|^2, \label{Ntensor2a} \\ 
&& \textrm{spin $\frac{1}{2}$}: ~W_1=\frac{Q^2}{4 m_T^2} (F_1+F_2)^2,~W_2=F_1^2+\frac{Q^2}{4m_T^2}F_2^2. \label{Ntensor2b}
\eea
 $F(Q^2)$ is the charge form factor. $F_1$ and $F_2$ are the Dirac and Pauli form factors respectively.
For nucleon, the electromagnetic form factors are familiar to us and are given  in Appendix B.

Using (\ref{Ltensor2b}), (\ref{Ntensor2a}) and (\ref{Ntensor2b}), 
the cross section for scattering $e + T \to \tau +T$  induced by (\ref{ETauGammaV})  can be found as
\bea
&& \frac{d \sigma}{d Q^2}=\frac{2 \pi Z^2 \alpha^2}{E^2 Q^4}(|c_L|^2+|c_R|^2)  \frac{|F(Q^2)|^2}{m_T^2}  \big[ 4 Q^2 (P\cdot k)^2 \nnb \\
&& ~~~+(Q^2+m^2_\tau)(P\cdot q)^2-(Q^2+m^2_\tau)(4P\cdot q P\cdot k+m^2_T m^2_\tau) \big ]  \label{Xsec2a}
\eea
for T of spin 0 , and 
\bea
\frac{d \sigma}{d Q^2} &&=\frac{2 \pi Z^2 \alpha^2}{E^2 Q^4}(|c_L|^2+|c_R|^2) \bigg \{ W_1(Q^2+m_\tau^2)(2 m^2_\tau -Q^2)  
+\frac{W_2}{m_T^2}\bigg[ 4 Q^2 (P\cdot k)^2  \nnb \\
&& +(Q^2+m^2_\tau)(P\cdot q)^2-(Q^2+m^2_\tau)(4P\cdot q P\cdot k+m^2_T m^2_\tau) \bigg ]  
\bigg \}\label{Xsec2b}
\eea
for T of spin $\frac{1}{2}$.  
Z  in (\ref{Xsec2a}) and (\ref{Xsec2b}) is the atomic number of nucleus for scattering with nucleus.

\begin{figure}[tb]
\begin{center}
\begin{tabular}{cc}
\includegraphics[scale=1,width=9cm]{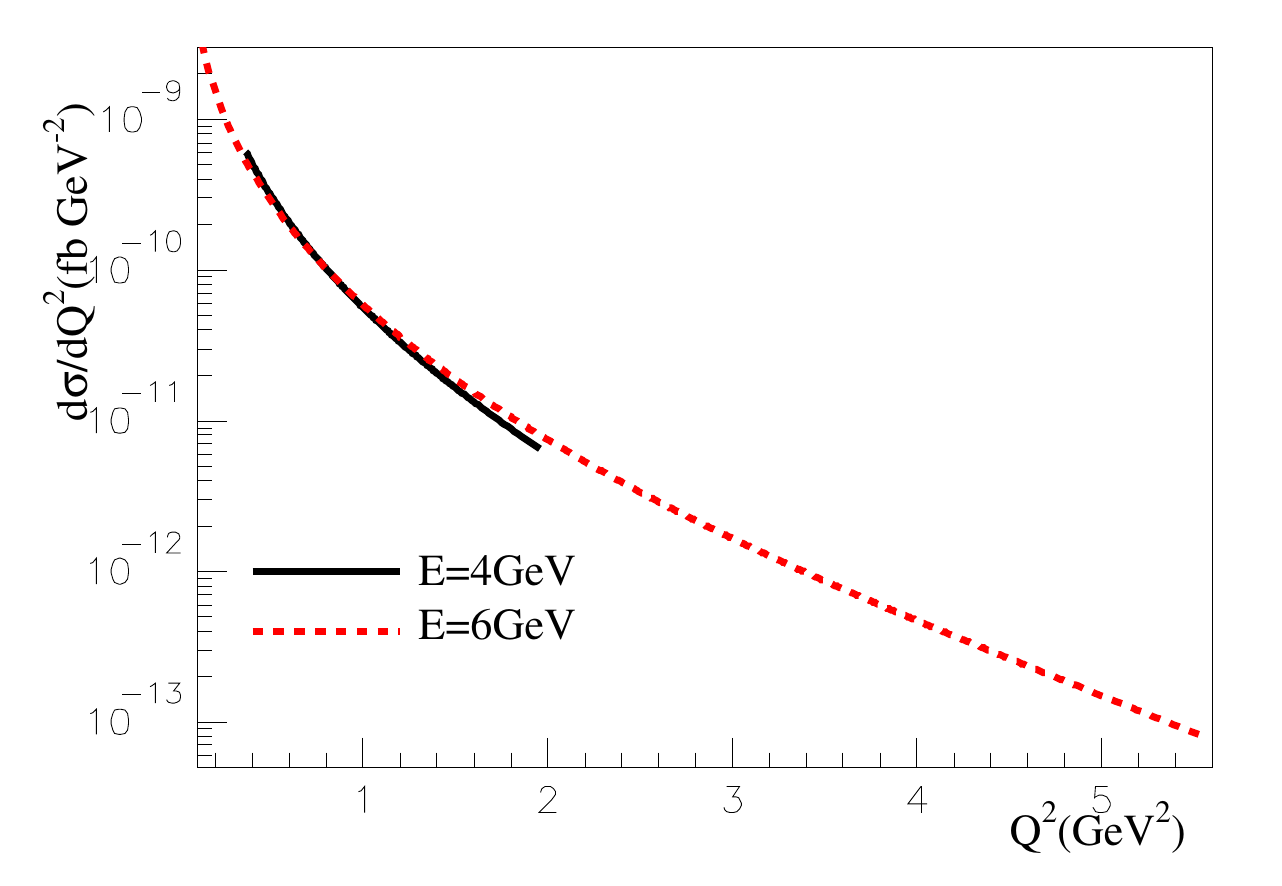}
\end{tabular}
\end{center}
\caption{Differential cross section of  $e +p \to \tau +p$ versus $Q^2$ induced by lepton flavor violating
$e\tau \gamma$ interaction.}
\label{Xsec1Q2}
\end{figure}

For elastic scattering with proton we just need to take $m_T=m_p$ and $Z=1$ in Eq. (\ref{Xsec2b}).
In Fig. \ref{Xsec1Q2} we show numerically the differential cross section versus $Q^2$.  
In the plot we take $|c_L|^2+|c_R|^2$ equal to the upper bound given in (\ref{constraint2}).
One can see that the cross section is very small. 
It is suppressed not only by $|c_L|^2+|c_R|^2$ which is of order $10^{-8}$ in terms of $G_F^2 m^2_\tau$, 
the square of the Fermi constant times the mass of $\tau$ lepton, 
but also suppressed by the appearance of $\alpha^2$ which gives a further suppression of order $10^{-4}$.
The total cross sections for curves of $E=4$ GeV and $E=6$ GeV in Fig.  \ref{Xsec1Q2} are
about $1.5\times 10^{-10}$ fb and $5.4\times 10^{-10}$ fb respectively.
So it's very hard to probe the lepton flavor violating $e \tau \gamma$ interaction vertex to a good sensitivity
using $e+N \to \tau+N$  scattering with energy of a few GeV.

One way to enhance the sensitivity is to use heavy nuclei so that $Z^2 \sim 10-100$.
However,  the size of nuclei is of fm scale. As a consequence, the nuclear form factors in elastic scattering,
which reflect how much nuclei are like point particles, drop down quickly at the scale of a few fm$^{-1}$.
For example, the form factors of light nuclei drop down by two to three orders of magnitude at scale 
3-4 fm$^{-1}$\cite{formfactor01}  which corresponds to  $Q^2 \approx 0.3-0.6 $ GeV$^2$. 
So there is only a very small region of phase space, namely region with $Q^2 \ll 1$GeV$^2$,  that the cross section
can potentially be enhanced by $Z^2$.  Moreover, we have learned from Table \ref{Q2min} that 
$Q^2$ is limited by the appearance of the massive $\tau$ lepton in the final state and 
is not allowed to be arbitrarily close to zero for a fixed energy. This is, in particular, for the cases of small energies.
So we do not expect that using heavy nuclei would help much in improving
the sensitivity on the lepton flavor violating $e \tau \gamma$ interaction.

The quasi-elastic scattering with nucleus is basically considered as the elastic scattering with nucleons in the
nucleus, as explained in the last section. So this process is not going to improve the sensitivity on
the lepton flavor violating $e \tau \gamma$ interaction vertex either,  for exact the same reason
for elastic scattering with free protons discussed above.

 \vskip 0.5cm
{\bf Probing $e\tau Z$ interation}

As another example of lepton flavor violating interaction,  $e-\tau$ conversion induced by the interaction
 with a Z boson is also important.
 In this section we study the  sensitivity of lepton flavor violating ES and QE scattering processes
 on the $e \tau Z$ interaction vertex.

The lepton flavor violating $e\tau Z$ interaction vertex can be written as
\bea
\Delta L=\frac{g}{\cos\theta_w} [C_L Z^\mu {\bar \tau}_L\gamma_\mu e_L+C_R Z^\mu {\bar \tau}_R\gamma_\mu e_R +h.c.]
\label{ETauZV}
\eea
where $C_{L,R}$ are the coupling strength, normalized in unit of $g/\cos\theta_W$, the SM coupling of Z boson with the associated current.
The decay rate of the $Z\to e^{\pm}\tau^{\mp}$ decay can be calculated using (\ref{ETauZV})
\bea
\Gamma(Z\to e^{\pm}\tau^{\mp}) =\frac{g^2 m_Z}{12\pi \cos^2\theta_w} (|C_L|^2+|C_R|^2),  \label{Decayrate}
\eea 
where $m_Z$ is the mass of Z boson. 
Comparing with the SM coupling for $\nu$, i.e. $C_\nu =1/2$,  we can find that
\bea
\frac{|C_L|^2+|C_R|^2}{|C_\nu|^2}=\frac{3}{2}\frac{Br(Z\to e^{\pm}\tau^{\mp})}{Br(Z\to \textrm{invisible})} \nnb
\eea
For the present upper bound Br$(Z\to e^{\pm}\tau^{\mp})< 9.8 \times 10^{-6}$\cite{pdg, akers}, we find
\bea
|C_L|^2+|C_R|^2 < 7.35 \times 10^{-5} |C_\nu|^2. \label{Bound1}
\eea

Another important constraint on (\ref{ETauZV}) comes from the $\tau \to 3 e$ decay. One can find that
\bea
Br(\tau \to 3 e) && =\tau_\tau \frac{G_F^2 m_\tau^5}{96 \pi^3}(|C_L|^2+|C_R|^2)[ (-\frac{1}{2}+\sin^2\theta_W)^2+\sin^4\theta_W] \nnb \\
&& =0.125~ \tau_\tau \frac{G_F^2 m_\tau^5}{96 \pi^3}(|C_L|^2+|C_R|^2),  \label{Decayrate1}
\eea
where G$_F$ is the Fermi constant.
For the present upper bound Br$(\tau \to 3 e) < 2.7 \times 10^{-8}$\cite{pdg}, we get
\bea
|C_L|^2+|C_R|^2 < 6 \times 10^{-7} \label{Bound2}
\eea
(\ref{Bound2}) is stronger than (\ref{Bound1}), the bound from $Z\to e^\pm \tau^\mp$ decay,  by about a factor 30.
Other constraints, such as those from $\tau \to e+\pi^+ + \pi^-$,  $\tau \to e+\mu^+ + \mu^-$ etc,  are of the same order of magnitude as
the $\tau \to 3e$ constraint and we are not going to elaborate on all of them in this article.

For one Z exchange, one can write the cross section of $e + T \to \tau +T'$ scattering as follows

\bea
\frac{d \sigma}{d Q^2}=\frac{G_F^2}{32\pi E^2} W_{\mu \nu} L^{\mu \nu} \label{Xsec}
\eea
where $W_{\mu \nu}$ is the nuclear(hadronic) tensor and $L^{\mu \nu}$ is the leptonic tensor.
In general $W_{\mu \nu}$ can be written as
\bea
W_{\mu \nu}&&=-\eta_{\mu \nu} W_1+\frac{P_\mu P_\nu}{m_T^2} W_2+
\frac{i \epsilon_{\mu \nu \rho \sigma} P^\rho q^\sigma }{2 m^2_T} W_3 \nnb \\
&&+\frac{q_\mu q_\nu}{m^2_T} W_4 +\frac{P_\mu q_\nu +P_\nu q_\mu}{2 m^2_T} W_5
+i\frac{P_\mu q_\nu -P_\nu q_\mu}{2 m^2_T} W_6. \label{NTensor}
\eea
This nuclear tensor is common to one Z exchange processes with the same $T'$  irrespective of the leptonic part.
In particular, $W_{1,2,3}$ appear in the $\nu+T \to \nu +T'$ scattering and can be measured using the scattering with neutrino.

In our convention,  the leptonic tensor $L^{\mu \nu}$  of $e +T \to \tau +T'$  scattering for unpolarized electron or positron
beam due to the $e\tau Z$ interaction (\ref{ETauZV}) is 
\bea
L^{\mu \nu} && =16[ (k^\mu k'^\nu+k^\nu k'^\mu -k\cdot k' \eta^{\mu \nu})(|C_L|^2+|C_R|^2)\nnb \\
&& +i \varepsilon^{\mu \nu \rho \sigma} k_\rho q_\sigma (|C_L|^2-|C_R|^2 )  ].\label{Ltensora}
\eea 
As a comparison, the leptonic tensor for  $\nu+T \to \nu +T'$ is
\bea
L^{\mu \nu}=8 (k^\mu k'^\nu+k^\nu k'^\mu -k\cdot k' \eta^{\mu \nu} +i \varepsilon^{\mu \nu \rho \sigma} k_\rho q_\sigma) \label{Ltensorb}
\eea 
Inserting (\ref{Ltensorb}) into the expression of cross section gives an expression depending only on $W_{1,2,3}$ .
Contributions of $W_{4,5}$ vanishes in the limit of massless neutrino. For $e+T \to \tau +T'$ scattering, the mass of the $\tau$ lepton
can not be neglected in general and terms with $W_{4,5}$ should be included.  However, there is no experimental data in principle
to extract $W_{4,5}$ as for $W_{1,2,3}$ in the scattering with neutrino. We should rely on the theoretical prediction. 
Fortunately, for ES  scattering with nucleon and the QE scattering, $W_{1,2,3,4,5}$ can all be expressed 
using the form factors of nucleons which are known quite well. They are discussed in Appendix C.

Now we consider the elastic scattering with nucleon $e+N \to \tau +N$.
For unpolarized electron or positron beam, the cross section of $e^\mp +N \to \tau^\mp + N$ elastic scattering is
\bea
\frac{d \sigma}{d Q^2} &&=\frac{G_F^2}{2 \pi E^2} \bigg \{ (|C_L|^2+|C_R|^2) \big [ (m^2_\tau +Q^2) (W_1-\frac{1}{2}W_2)  \nnb \\
&& +\frac{P\cdot k}{m_N^2}(2 P\cdot k-Q^2) W_2  +\frac{1}{2} (m^2_\tau+Q^2)\frac{m^2_\tau}{m_N^2} W_4
-\frac{m^2_\tau}{m_N^2} P\cdot k W_5 \big]   \nnb \\ 
&& \pm (|C_L|^2-|C_R|^2) \frac{Q^2}{4 m_N^2}(4 P\cdot k -Q^2-m^2_\tau) W_3  \bigg\}  \label{Xsec1}
\eea
$W_i$($i=1,2,3,4,5$) are given by (\ref{W1}),  (\ref{W2}),  (\ref{W3}),  (\ref{W4}) and  (\ref{W5}).

 \begin{figure}[tb]
\begin{center}
\begin{tabular}{cc}
\includegraphics[scale=1,width=8cm]{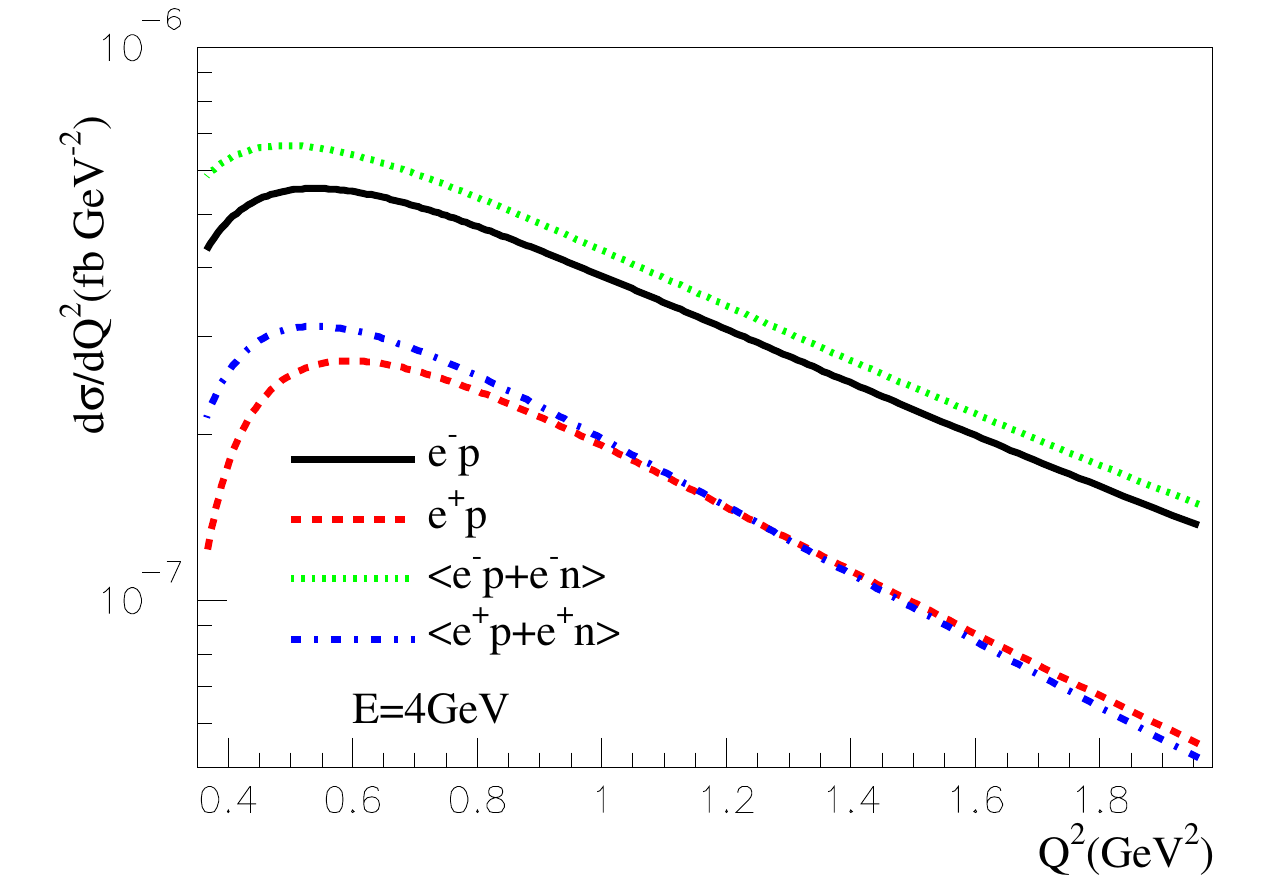}
\includegraphics[scale=1,width=8cm]{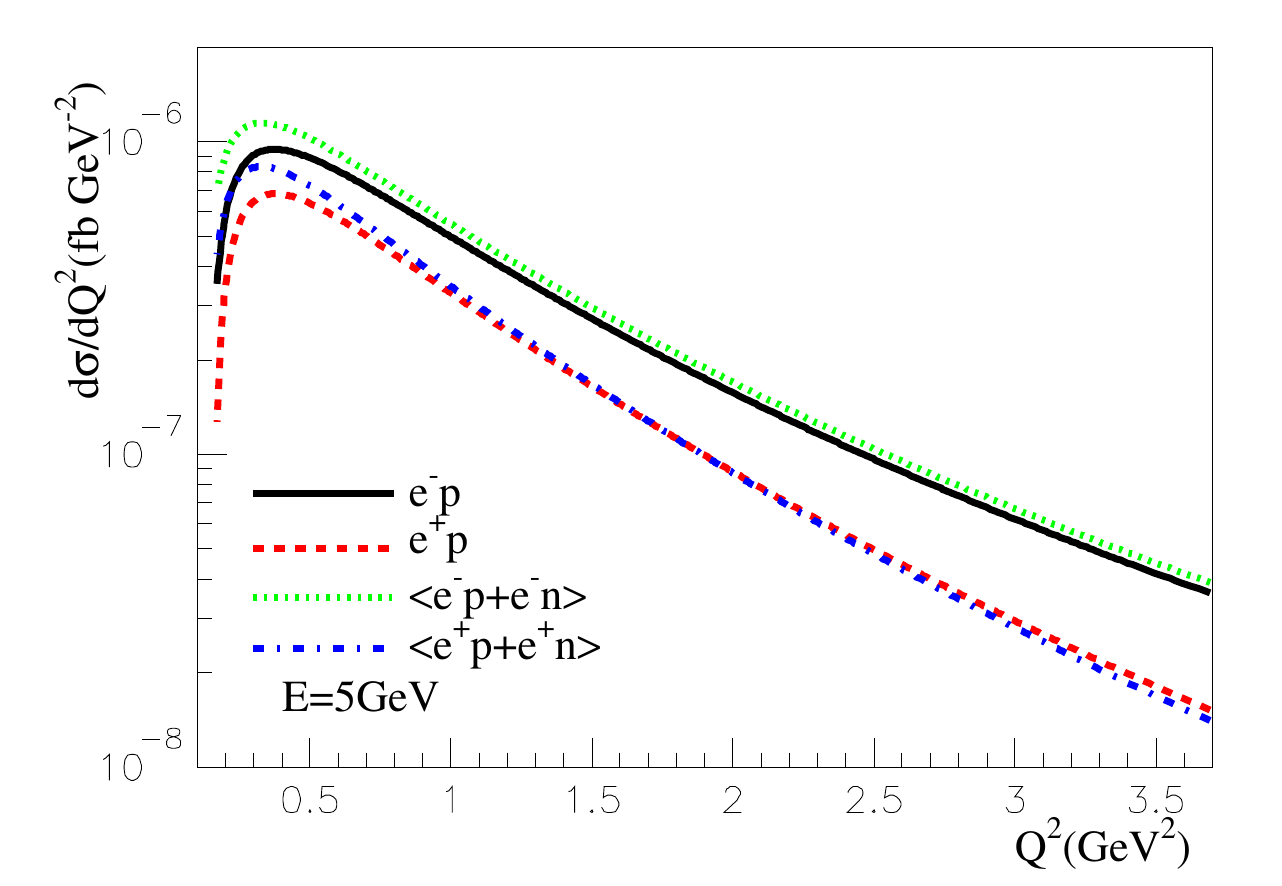}
\\
\includegraphics[scale=1,width=8cm]{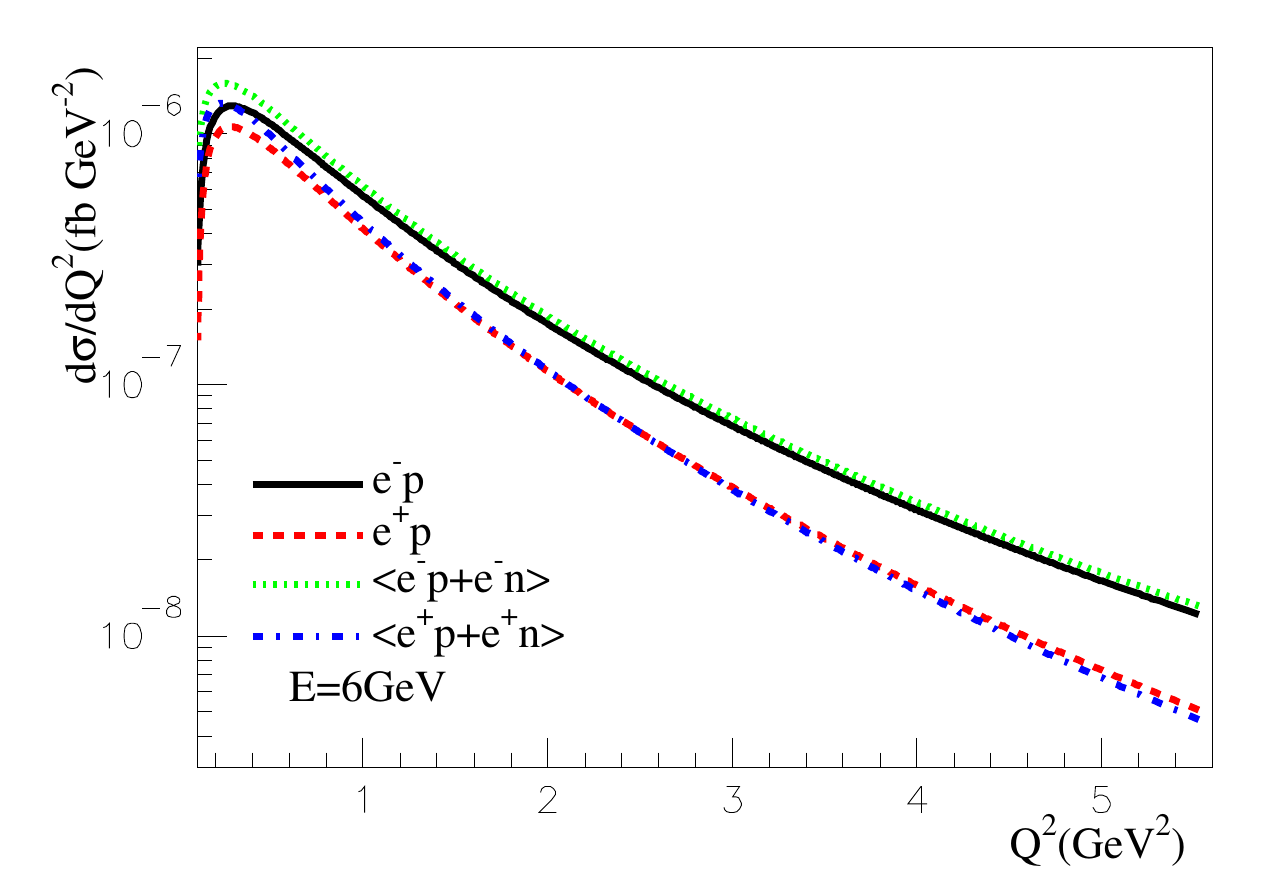}
\includegraphics[scale=1,width=8cm]{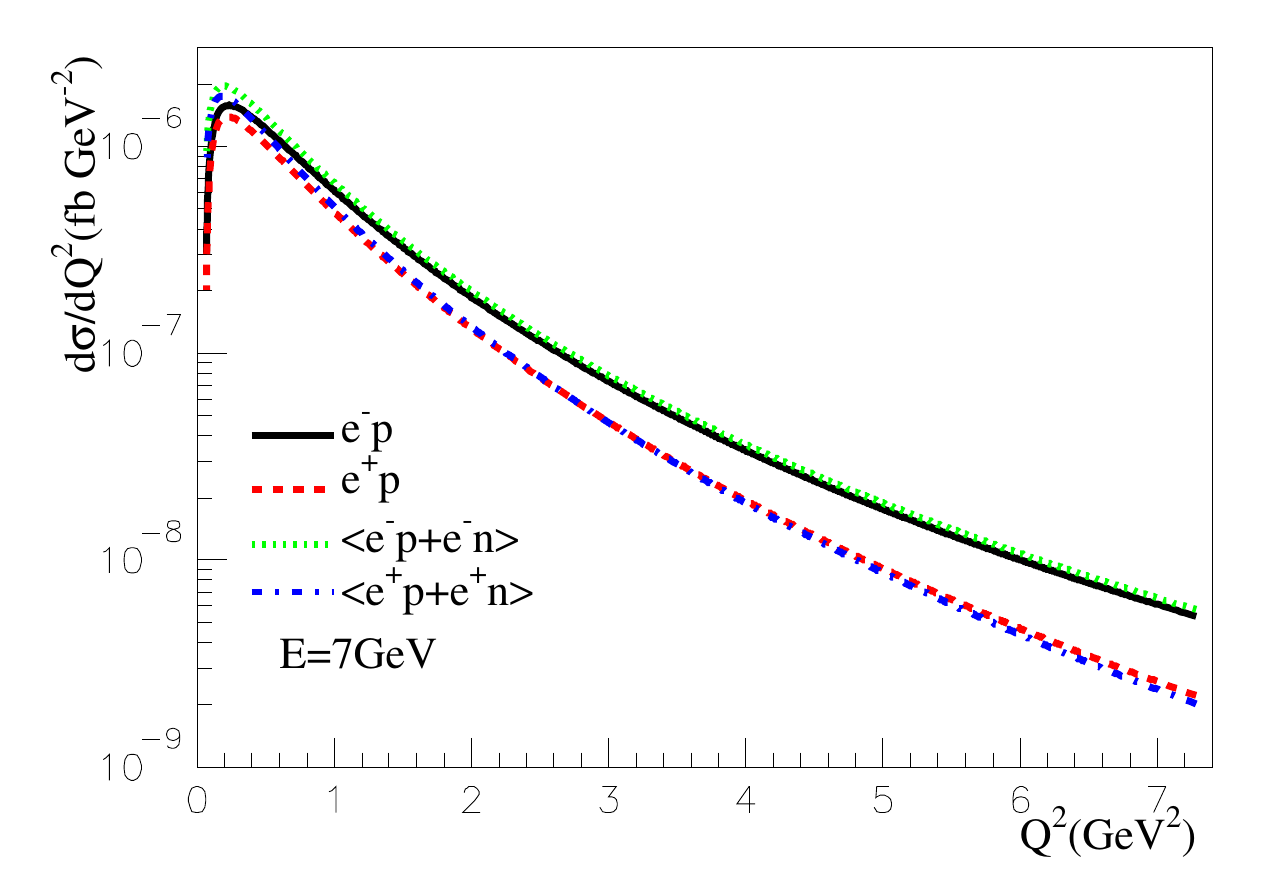}
%\\
%\includegraphics[scale=1,width=7.5cm]{figure1e}
%\includegraphics[scale=1,width=7.5cm]{figure1f}
\end{tabular}
\end{center}
\caption{Differential cross section of  $e +N \to \tau +N$ versus $Q^2$ induced by lepton flavor violating $e\tau Z$ interaction.  
$e^\pm p$ lines are for  scattering with proton
and $\langle e^\pm p +e^\pm n \rangle$ lines are the averages of cross sections of  the scattering with proton and neutron.}
\label{Xsec2Q2}
\end{figure}

In Fig. \ref{Xsec2Q2}  we can see numerically the differential cross section of $e^\mp +N \to \tau^\mp + N$ elastic scattering
versus $Q^2$ induced by $e\tau Z$ interaction in (\ref{ETauZV}). 
In these plots we have set $C_R=0$ and the upper bound value in (\ref{Bound2}) has been used for $C_L$.
In this case with $C_R=0$,  the cross section for electron beam is larger than the corresponding cross section for positron beam. 
If $C_L=0$,  the cross section for positron beam is larger.
The curves for $\langle ep +en\rangle$ are the averages of cross sections for proton and neutron targets. They are the average
cross sections of scattering with nucleon for isoscalar target. 
We can see that the cross section is not that small as for the $e-\tau$ conversion induced by interaction with photon.
For energy of $4-7$ GeV as shown in Fig. \ref{Xsec2Q2}, the total cross sections 
for $e^- p$,  $e^+ p$, $\langle e^- p+e^- n \rangle$ and $\langle e^+ p+e^+ n \rangle$
vary in ranges $(0.63-1.6)\times 10^{-6}$ fb, $(0.28-1.2)\times 10^{-6}$ fb,  $(0.71-1.8)\times 10^{-6}$ fb, and $(0.30-1.4)\times 10^{-6}$ fb
separately. So for this process to reach the sensitivity of $\tau \to 3e$ decay, the luminosity of experiment should reach
1000 ab$^{-1}$. On the other hand, to reach the sensitivity of $Z \to e^\pm \tau^\mp$, as shown in (\ref{Bound1}),
the luminosity should reach around 30 ab$^{-1}$.

For quasi-elastic scattering with nuclei,  nucleons are kicked out of the nuclei by elastic scattering with
the initial electron or positron. The cross section can be calculated using that of  scattering with free nucleon $e^\mp +N \to \tau^\mp +N$.
Nuclear corrections to the free nucleon approximation have been
studied, for example, for the quasi-elastic scattering of neutrino with nuclei\cite{freeNApp}, which is called neutral current elastic(NCE) 
scattering in the neutrino physics community. The results show that for energy of a few GeV, and for $Q^2$ not very close to zero,
the free nucleon approximation works perfectly. So it's fairly good to adopt this approximation when we  estimate the sensitivity
of probing lepton flavor violating $e \tau Z$ interaction. It's certainly true that we should take these nuclear corrections
into account if we want more detailed understandings. But this detailed work should be done for those matter 
selected to use as targets in experiments. We are not going to elaborate on this topic in this article and leave it to future research.

The signals of the lepton flavor violating processes discussed above are simple and easy to distinguish. 
In addition to a $\tau$ lepton, final particles of the processes include a free nucleon in the case of elastic scattering, 
or several pieces of the broken nuclei in the case of quasi-elastic scattering.
If the energy of the $\tau$ lepton can be reconstructed to a good precision, there should be no missing energy in these final products
except that arising from the $\tau$ decay.  

Processes with production of heavy hadrons by strong interaction, e.g. $e+N\to e +N+D^+ +D^-$ or $e+N \to e +N+D_s^+ +D_s^-$,  
can potentially give $\tau$ lepton in decays of these hadrons. 
So in principle they can mimic the lepton flavor violating process $e +N \to \tau +N$
discussed in this article if the scattered electron or positron and the decay products other than $\tau$ lepton all escape the detection of detector. 
A great virtue of considering $e +N \to \tau +N$ scattering at a few GeV is that
we can avoid these potential backgrounds. One can show  that a threshold condition similar to (\ref{kinematics05})
holds for  the production of a pair of D mesons or D$_s$ mesons. Consequently, the energy threshold is more than 4 times 
the threshold for  $e +N \to \tau +N$ scattering shown in Table \ref{Threshold}.  By comparing with Table \ref{Threshold}, 
we can see that this potential background would disappear if considering experiments with beam energy $< 8$ GeV. 
For the same reason, process with production of a pair of $\tau$ lepton by neutral current weak interaction can
also be avoided if considering experiments with beam energy $< 8$ GeV. 

A relevant process for background consideration is the process with production 
of a pair of $(\tau^-, {\bar \nu}_\tau)$ or $(\tau^+, \nu_\tau)$ by charged current weak interaction. 
The charge in the target part can be balanced by emitting a soft pion.  
This pion, which should escape the detection by the detector,  should carry a small amount of energy, 
so that this undetected particle does not affect much the energy budget of the scattering process detected in detector. 
 So for $e^- +N \to \tau^- +N$ or
$e^+ +N \to \tau^+ +N$ scattering, the relevant background processes are  
$e^- +N \to e^- +N+ \tau^- +{\bar \nu}_\tau +\pi^+$ or $e^+ +N \to e^+ +N+ \tau^ + +\nu_\tau +\pi^-$ respectively.
If the scattering with the nucleon is through weak interaction, the total cross section with
the production of a $\tau$ lepton is doubly suppressed by the weak interaction, that is  it would be
proportional to $G_F^4 E^6$ which is suppressed by $G_F^2 E^4\sim 10^{-9}$  for energy of a few GeV, compared to
the usual weak interaction process. So this kind of contribution to background is negligible.
At leading order, the background process of $e^\mp$ scattering with nucleon is given by a photon exchange which gives a factor $\alpha^2$
in the cross section.  So this background process is proportional to $\alpha^2 G_F^2 E^2$ which is $\sim 10^{-4}$ fb for
energy of a few GeV.  Moreover, there are three extra particles in the background process compared to the
signal process, namely $e^\mp$, ${\bar \nu}_\tau$ or $\nu_\tau$,  and $\pi^\pm$. Their appearance  in the final
state contribute three phase space factors which altogether should give a suppression factor no less than $10^{-3}$. 
Finally, the scattered $e^\mp$ should also escape the detection by the detector and should also
carry a small amount of energy, similar to the consideration for soft pion. 
This means that to mimic the lepton flavor violating process $e^\mp +N \to \tau^\mp +N$,
final $e^\mp$ and $\pi^\pm$ of the background process are limited to a small region of phase space, 
e.g. $\lsim 10\%$ of the momentum region allowed by the kinematics.
So the cross sections of the background event production are further suppressed by these phase space considerations.
Taking all these into account, we expect that the background process is at most of order $10^{-9} G_F^2 E^2 $ which is
around $10^{-9}$ fb for energy of a few GeV.  It's several orders of magnitude smaller than the cross section of the
signal process presented in Fig.  \ref{Xsec2Q2}. 

Another kind of possible background process is through charged current interaction:
$e^- +N \to \nu_e +N+ \tau^- +{\bar \nu}_\tau $ or $e^+ +N \to {\bar \nu}_e +N+ \tau^ + +\nu_\tau $.
The cross sections of these processes are either proportional to $G^4_F E^6$ or $\alpha^2 G_F^2 E^2$. 
For reasons similar to the above discussions, we expect these processes also give negligible background events.
So background processes are negligible and there are quite a lot room for studying lepton flavor violating effect in $e+N \to \tau +N$ process. 
More detailed analysis of the background processes is not the topic of this article and we leave it to future works.

\vskip 0.5cm
{\bf Summary}

In summary, we have studied the possible lepton flavor violating effects in a simple process of
electron or positron scattering with nucleon,  $e + N \to \tau +N$,  at energy of a few GeV.  
This lepton flavor violating process $e + N \to \tau +N$ can be searched for by directing electron or positron beam 
on internal or fixed target. For scattering with hydrogen target, this process is the elastic scattering of electron or positron with
proton. For target of heavier nuclei, such as deuteron, helium, or others, this process can be studied in quasi-elastic scattering
of $e^\mp$ with nuclei.  The final particles in this quasi-elastic scattering are some nucleons or nuclei with no extra
hadron produced, i.e. just the broken pieces of the initial nuclei, plus a $\tau$ lepton. 
So the final states are quite simple to study in experiments.

We have considered, as two model independent examples,  lepton flavor violating $e \tau \gamma$ and $e \tau Z$  interactions
which are possible low energy effective interactions arising from physics beyond SM.  We have studied the effects of these interactions on
the process  $e + N \to \tau +N$ and calculated the cross section of this process induced
by these two kind of interactions.  We compare the sensitivity of other lepton flavor violating process, 
such as $\tau \to e \gamma$, $\tau\to 3e$, $Z \to e^\mp \tau^\pm$ etc, and the sensitivity of the process
$e + N \to \tau +N$ on the lepton flavor violating interactions.  We have shown that the lepton flavor
violating process $e + N \to \tau +N$ is not as sensitive to lepton flavor violating $e \tau \gamma$ interaction,
as to $e \tau Z$ interaction. For the later, a target experiment with luminosity $\gsim 30$ ab$^{-1}$ can
give a better sensitivity on the $e \tau Z$ interaction vertex than $Z \to e^\mp \tau^\pm$.  To be better
than the sensitivity of  $\tau\to 3e$ decay, the luminosity is required to reach 1000 ab$^{-1}$.

 For simplicity, we have concentrated on this lepton flavor violating process at GeV scale
which can be studied by GeV scale $e^\mp$ accelerator facilities at various places in the world.  
Another reason of considering GeV scale lepton flavor violating effect is that at this energy scale the elastic scattering or quasi-elastic
scattering is the dominant process and possible background from other channels, which could mimic the lepton
flavor violating signal, could be significantly suppressed.  So, unlike the processes at higher energy, the background
study could be significantly simplified at GeV scale.
Note that similar process with muon beam if it's available in experiment, i.e. $\mu + N \to \tau +N$, 
can also be considered and formalism discussed in this article can be similarly applied to 
elastic or quasi-elastic scattering of muon with target.
Needless to say,  this process provides an alternative  way to search for lepton flavor violating effect associated with $\tau$ lepton.
It is particularly interesting because GeV scale electron or positron beams
are available in a number of laboratories in the world, while it's hard to prepare the source of $\tau$ lepton and study the lepton
flavor violating effect in processes with $\tau$ lepton in initial state.

\acknowledgments
This work is supported by National Science Foundation of
 China(NSFC), grant No.11135009, No. 11375065.  W. Liao
 would like to thank Haibo Li for very helpful discussions on internal/fixed
 target experiment.

\vskip 1cm
%\appendix{Appendix A}
{\bf Appendix A}

Using (\ref{kinematics01}) and (\ref{kinematics02}),  we can find for $T'=T$ 
\bea
2 k\cdot k'-m^2_\tau=2 m_T (E-E') .\label{kinem01}
\eea
Using $\theta_{k'}$, the angle between ${\vec k}$ and ${\vec k}'$,  (\ref{kinem01}) can be rewritten as
\bea
2 EE'-2 E |{\vec k}'|\cos\theta_{k'} -m^2_\tau=2 m_T(E-E'), \label{kinem02}
\eea
or
\bea
(2 E+ 2m_T) E'-(2 m_TE +m^2_\tau)= 2 E |{\vec k}'|\cos\theta_{k'}, \label{kinem02a}
\eea
where $E=|{\vec k}|$ has been used and $|{\vec k}'|=\sqrt{E'^2-m^2_\tau}$.  Making a square of (\ref{kinem02a}) we get an equation for $E'$
which can be solved as
\bea
E'=\frac{(E+m_T)(m^2_\tau+2 m_T E)+E\cos\theta_{k'} \sqrt{A}}{2(E^2\sin^2 \theta_{k'}+2 E m_T+m^2_T)}, \label{kinem03}
\eea
where
\bea
A=(m^2_\tau+2 m_T E)^2-  4 m_\tau^2  (E^2 \sin^2 \theta_k'+2 E m_T+m^2_T).  \label{kinem04}
\eea
The condition to have a real solution of $E'$ is  $A \ge 0$. 
So we get a condition for the scattering angle $\theta_{k'}$ for a fixed initial energy $E$:
\bea
(m^2_\tau+2 m_T E)^2-  4 m_\tau^2  ( 2 E m_T+m^2_T) \ge 4 m_\tau^2 E^2 \sin^2 \theta_{k'}. \label{kinem05}
\eea
Taking the limit $m_\tau=0$ in (\ref{kinem03}) we can find that the solution becomes
\bea
E'=\frac{m_T E}{E+m_T-E\cos\theta_{k'}}=\frac{E}{1+\frac{E}{m_T}(1-\cos\theta_{k'})}, \label{kinem03a}
\eea
which is the familiar formula of elastic scattering in massless limit.

For $T'\ne T$ we can find from  (\ref{kinematics01}) and (\ref{kinematics02}) that 
\bea
2 k\cdot k'-m^2_S=2 m_T (E-E') \label{kinem06}
\eea
where 
\bea
m^2_S= m^2_\tau-(m^2_{T'}-m^2_T). \label{kinem07}
\eea
The solution of $E'$ can be obtained with a similar procedure 
\bea
E'=\frac{(E+m_T)(m^2_S+2 m_T E)+E\cos\theta_{k'} \sqrt{A'}}{2(E^2\sin^2 \theta_{k'}+2 E m_T+m^2_T)}, \label{kinem08}
\eea
where
\bea
A'=(m^2_S+2 m_T E)^2-  4 m_\tau^2  (E^2 \sin^2 \theta_{k'}+2 E m_T+m^2_T).  \label{kinem09}
\eea
The condition (\ref{kinem05}) becomes
\bea
(m^2_S+2 m_T E)^2-  4 m_\tau^2  ( 2 E m_T+m^2_T) \ge 4 m_\tau^2 E^2 \sin^2 \theta_{k'} \label{kinem10}
\eea
Applying (\ref{kinem10}) to $\theta_{k'}=0$ gives (\ref{kinematics03}), the threshold condition of the energy.

\vskip 1cm
{\bf Appendix B}

For elastic scattering $e + T \to \tau +T$ induced by a photon exchange, 
the electromagnetic form factors of nucleus or nucleon are introduced as follows.
For spin 0 nuclear target, it is given in
\bea
\langle P+q | J^\mu_{EM}| P \rangle =F(Q^2) (P^\mu -\frac{P\cdot q}{q^2} q^\mu).  \label{formfactors01}
\eea
For spin $\frac{1}{2}$ target, it is given in
\bea
\langle P+q, s' | J^\mu_{EM}| P,s  \rangle = {\bar u}_{s'}(P')
\big[ F_1(Q^2)\gamma^\mu+i\frac{F_2(Q^2)}{2 m_T} \sigma^{\mu \rho} q_\rho \big ] u_s(P),
\label{formfactors02}
\eea
where $P'=P+q$,  $s'$ and $s$ are spin indices. $J^\mu_{EM}$ is the electromagnetic current.
$F_1$ and $F_2$ are Dirac and Pauli form factors.

For nucleon target, we take $m_T=m_N$ in (\ref{formfactors02}) where $m_N= m_p\approx m_n$.
The corresponding form factors are the nucleon form factors $F^N_i$($i=1,2$) , and
\bea
F_i^N = \left \{ \begin{matrix}   F^p_i, ~\textrm{N=proton} \cr
 F^n_i, ~\textrm{N=neutron}  \end{matrix} \right.
\eea
These form factors satisfy: 
\bea
&& F^p_1(Q^2=0)=1 , ~~F^p_2(Q^2=0)=\kappa_p \\
&& F^n_1(Q^2=0)=0,  ~~F^p_2(Q^2=0)=\kappa_p, 
\eea
where $\kappa_p$ and $\kappa_n$ are anomalous magnetic moments in unit of $e/2m_N$ of proton and neutron respectively.
% $\kappa_p=1.793$, $\kappa_n=-1.91$.

$F^{p,n}_i$ can be expressed  using Sachs form factors $G_E$ and $G_M$
\bea
F_1^{p,n}(Q^2) &&=\frac{G_E^{p,n}(Q^2)+\frac{Q^2}{4 m_N^2} G_M^{p,n}(Q^2)}{1+\frac{Q^2}{4 m_N^2}}, \label{formfactors06a} \\
F_2^{p,n}(Q^2) && =\frac{G_M^{p,n}(Q^2)- G_E^{p,n}(Q^2)}{1+\frac{Q^2}{4 m_N^2}}. \label{formfactors06b}
\eea
In dipole approximation, $G_E$ and $G_M$ are expressed as
\bea
G^{p,n}_{E,M}(Q^2)=\frac{G^{p,n}_{E,M}(Q^2=0)}{(1+Q^2/M_V^2)^2}, \label{formfactors06c}
\eea
where the vector mass is $M_V=0.843$ GeV, and 
\bea
&& G^p_E(Q^2=0)=1,~~G^p_M(Q^2=0)=\mu_p, \label{formfactors06d} \\
&& G^n_E(Q^2=0)=0,~~G^n_M(Q^2=0)=\mu_n .\label{formfactors06e}
\eea
where $\mu_N=2.793$ and $ \mu_n=-1.91$ are the total magnetic moment  in unit of $e/2m_N$ of proton and neutron respectively.
In this article we adopt the dipole approximation of these electromagnetic form factors of nucleons.
More sophisticated and complicated approximations can be found in literatures~\cite{mmd,formfactorfit} 
which can be used for more accurate calculation. In this article, we do not use them in our calculation and in our
estimate of the sensitivities on lepton flavor violating interactions.

\vskip 1cm
{\bf Appendix C}

Using the lepton flavor violating effective coupling (\ref{ETauZV}),  a neutral current effective interaction with quark
can be induced at low energy by a Z exchange.
For the scattering with nucleon, $e +N \to \tau +N$,  the relevant interaction is
\bea
\Delta L &&=\frac{2G_F}{\sqrt{2}} [ C_L {\bar \tau}\gamma_\mu (1-\gamma_5) e+ C_R {\bar \tau}\gamma_\mu (1+\gamma_5) e ] \nnb \\
&& \times \sum_i \big[ {\bar q}_i \frac{1}{2}T^3 \gamma^\mu (1-\gamma_5) q_i-2 \sin^2\theta_W J^\mu_{EM} \big], \label{EffCoupling1}
\eea
where $\theta_W$ is the weak mixing angle, $q=$u or d quark,  $T^3$ the isospin operator, $J^\mu_{EM}$ the electromagnetic current
for nucleon
\bea
J^\mu_{EM}=\sum_i Q_i {\bar q}_i \gamma^\mu q_i  =\frac{2}{3} {\bar u} \gamma^\mu u-\frac{1}{3} {\bar d} \gamma^\mu d.
\eea
For simplicity we have neglected possible contribution of strange sea quark~\cite{abm}.
The quark part of the current in (\ref{EffCoupling1}) can be decomposed into isoscalar and isovector part and (\ref{EffCoupling1}) can be re-written as
\bea
\Delta L &&=\frac{2G_F}{\sqrt{2}} [ C_L {\bar \tau}\gamma_\mu (1-\gamma_5) e+ C_R {\bar \tau}\gamma_\mu (1+\gamma_5) e ] \nnb \\
&& \times \big [   x V^\mu_3+y V^\mu_0 +\gamma  A^\mu_3+\delta A^\mu_0  \big], \label{EffCoupling2}
\eea
where
\bea
&& V^\mu_3=\frac{1}{2}({\bar u} \gamma^\mu u-{\bar d} \gamma^\mu d),~~
V^\mu_0=\frac{1}{6}({\bar u} \gamma^\mu u+{\bar d} \gamma^\mu d),   \label{Decomp1a}\\
&& A^\mu_3=\frac{1}{2}({\bar u} \gamma^\mu \gamma_5 u-{\bar d} \gamma^\mu \gamma_5 d),~~
A^\mu_0=\frac{1}{2}({\bar u} \gamma^\mu \gamma_5 u+{\bar d} \gamma^\mu \gamma_5 d),  \label{Decomp1b}
\eea
and for this coupling through Z exchange
\bea
x=1-2\sin^2\theta_W, ~y=-2\sin^2\theta_W,~\gamma=-1,~\delta =0 . \label{Decomp2}
\eea
 $V^\mu_0$ and $A^\mu_0$ are isoscalar currents, $V^\mu_3$ and $A^\mu_3$ are isovector currents.
Using (\ref{EffCoupling2}),  the matrix element between nucleon state is 
\bea
\langle N(P') | x V^\mu_3+y V^\mu_0 +\gamma  A^\mu_3+\delta A^\mu_0 | N(P) \rangle  \label{HadMatrixE1}
\eea

The vector part of (\ref{HadMatrixE1}) can be expressed using the electromagnetic form factors
$F^N_{1,2}$ of nucleon as follows. Since $J^\mu_{EM}$ can be rewritten as a sum of isovector and isoscalar component as
\bea
J^\mu_{EM}=V^\mu_0+V^\mu_3,  \label{JEM}
\eea
(\ref{formfactors02})  for nucleon can be re-written as
\bea
&& \langle N(P'), s' | J^\mu_{EM} | N(P), s \rangle \nnb \\
&& ={\bar u}_{s'}(P') \big[ (F^s_1+F^v_1\tau^3) \gamma^\mu 
+\frac{i}{2m_N} \sigma^{\mu \rho} q_\rho (F^s_2+F^v_2\tau^3) \big] u_s(P),  \label{formfactors04}
\eea
where $\tau^3$ takes value $+1$ for N=proton and takes value $-1$ for N=neutron.
$F^s_i$ and $F^v_i$ are isoscalar and isovector part of the form factors and are given by
the matrix elements of $V^\mu_0$ and $V^\mu_3$ respectively.
So we have
\bea
&& F_i^p=F^s_i+F^v_i, ~~i=1,2 \\
&& F_i^n= F^s_i-F^v_i~~i=1,2 ,  \label{formfactors05a}
\eea
or
\bea
F^s_i=\frac{1}{2}(F^p_i+F^n_i),~F^v_i= \frac{1}{2}(F^p_i-F^n_i). \label{formfactors05b}
\eea
Similar to (\ref{JEM}), the vector current in  (\ref{HadMatrixE1}) is also a linear combination of
isoscalar and isovector current $V^\mu_0$ and $V^\mu_3$.  So the vector part in (\ref{HadMatrixE1}) 
can be written as a linear combination of isoscalar and isovector contributions, similar to (\ref{formfactors04}).
So we get
\bea
&& \langle N(P') | x V^\mu_3+y V^\mu_0 | N(P) \rangle   \nnb \\
&& ={\bar u}_{s'}(P') \big[ 
F^z_1 \gamma^\mu +\frac{i}{2m_N} \sigma^{\mu \rho} q_\rho F^z_2 \big] u_s(P),
\label{HadMatrixE2}
\eea
where
\bea
F^z_i=y F^s_i+x F^v_i\tau^3, ~~i=1,2. \label{HadMatrixE2a}
\eea
$\tau^3$ takes value $+1$ for proton and $-1$ for neutron.
Inserting (\ref{formfactors05b}) into (\ref{HadMatrixE2a})  
we get an expression of matrix element  using the electromagnetic form factors $F^{p,n}_i$:
\bea
F^z_i=\frac{1}{2}(y +x \tau^3)F^p_i+ \frac{1}{2}(y -x \tau^3)F^n_i, ~~i=1,2. \label{HadMatrixE2b}
\eea

The axial part in (\ref{HadMatrixE1})  can be written into two terms ~\cite{abm} and 
similar to (\ref{HadMatrixE2}), each term  can be written as
a linear combination of isoscalar and isovector contributions.
\bea
&& \langle N(P') |  \gamma A^\mu_3+\delta  A^\mu_0 | N(P) \rangle   \nnb \\
&& ={\bar u}_{s'}(P') \big[ 
G^z_1 \gamma^\mu \gamma_5 +\frac{1}{2m_N}q^\mu\gamma_5  G^z_2 \big] u_s(P),
\label{HadMatrixE3}
\eea
where $G^z_i$ is a sum of isoscalar and isovector contributions
\bea
G^z_i=-\frac{\gamma}{2}  G^v_i\tau^3+\frac{\delta}{2} G^s_i=\frac{1}{2}  G^v_i\tau^3, ~~i=1,2. \label{HadMatrixE3a}
\eea
Again, $\tau^3$ takes value $+1$ for proton and $-1$ for neutron. 
Since $\delta=0$ for the scattering induced by Z exchange, only the isovector part $G^v_i$ give contribution in  (\ref{HadMatrixE3a}).
$G^v_1$ at $Q^2=0$ is known from the nucleon $\beta$ decay and for general $Q^2$ it can be taken as
\bea
G^v_1=g_A/(1+Q^2/M_A^2)^2 \label{formfactors07a}
\eea
in dipole approximation where $g_A=1.267$. For the axial nucleon mass $M_A$,  $M_A=1.39$ GeV is taken in this article~\cite{Miniboon}.  
$G^v_2$ is related to $G^v_1$ by the partial conservation of axial current(PCAC)
\bea
G^v_2=\frac{4 m_N^2}{m_\pi^2+Q^2}G^v_1 \label{formfactors07b}
\eea

Structure functions $W_i$ in (\ref{NTensor}) are expressed in terms of $F^z_i$ and $G^z_i$ as
\bea
W_1&&=\frac{Q^2}{4m_N^2}(F^z_1+F^z_2)^2+ (1+\frac{Q^2}{4m_N^2} )(G^z_1)^2  \label{W1} \\
 W_2 &&= (F^z_1)^2+\frac{Q^2}{4m_N^2} (F^z_2)^2+(G^z_1)^2 \label{W2}\\
 W_3 &&= 2 G^z_1 (F^z_1+F^z_2) \label{W3} \\
W_4 &&=\frac{1}{4}\big [ (F^z_1)^2+\frac{Q^2}{4m_N^2} (F^z_2)^2-(F^z_1+F^z_2)^2
+\frac{Q^2}{4m_N^2}(G^z_2)^2-2 G^z_1 G^z_2 \big] \label{W4}\\
W_5&&=(F^z_1)^2+\frac{Q^2}{4m_N^2} (F^z_2)^2+(G^z_1)^2\label{W5} \\
W_6&&=0. \label{W6}
\eea

For the neutral current  elastic scattering of neutrino or anti-neutrino with nucleon, $\nu({\bar \nu}) +N\to \nu({\bar \nu})+N$, the
cross section depends only on $W_{1,2,3}$, not on $W_{4,5}$.  By setting $m_\tau=0$ and
$C_R=0$ and normalizing to polarized beam which is appropriate for neutrino and anti-neutrino, 
one can recover  from (\ref{Xsec1})  the cross section for $\nu({\bar \nu}) +N\to \nu({\bar \nu})+N$ elastic scattering~\cite{LS,Miniboon, gklk}
\bea
\frac{d\sigma}{dQ^2}&& = \frac{G_F^2 Q^2}{2\pi E^2} 
\bigg \{ \frac{1}{2} \bigg[ W_1-\frac{1}{2}(1+\frac{Q^2}{4m^2_N}) W_2 \bigg] \nnb \\
&&~~~~\pm\frac{1}{8}\bigg( \frac{4E}{m_N}-\frac{Q^2}{m_N^2}\bigg) W_3
+ \frac{1}{16}\frac{m_N^2}{Q^2} \bigg(\frac{4E}{m_N}-\frac{Q^2}{m_N^2} \bigg)^2 W_2  \bigg \}. \label{Xsec1a}
\eea
(\ref{Xsec1a}) is also available by directly applying (\ref{Ltensorb}) into the cross section (\ref{Xsec}).

\end{document}